\documentclass[%
reprint,
superscriptaddress,
amsmath,amssymb,
aps,
prc,
]{revtex4-2}

\usepackage[detect-all]{siunitx}

\DeclareSIUnit{\weisskopfunit}{W.\hspace{1pt}u.}
\newcommand{\asymUnc}[4]{\ensuremath{#1\;^{+\;#2}_{-\;#3}\;\hspace{-0pt}\si{#4}}}
\newcommand{\bdown}{$B(E2)$$\downarrow$ }

\usepackage[version=4]{mhchem}
\usepackage{graphicx}
\usepackage{dcolumn}
\usepackage{bm}
\usepackage{hyperref}

\begin{document}
	
	\preprint{APS/123-QED}

    \title{Revised \textit{B}(\textit{E}\ce{2}; \ce{{2}^+_1} $\rightarrow$ \ce{{0}^+_1}) value in the semi-magic nucleus \ce{^{210}Pb}}   
	
	\author{C.~M.~Nickel}
	\email[Corresponding author: ]{cnickel@ikp.tu-darmstadt.de}
    \affiliation{
		Institute for Nuclear Physics, Dept. of Physics, Technische Universität Darmstadt, Schlossgartenstraße 9, D-64289 Darmstadt, Germany
	}
	\author{V.~Werner}
    \affiliation{
		Institute for Nuclear Physics, Dept. of Physics, Technische Universität Darmstadt, Schlossgartenstraße 9, D-64289 Darmstadt, Germany
	}
    \affiliation{
		Helmholtz Forschungsakademie Hessen für FAIR (HFHF), Campus Darmstadt, Schlossgartenstraße 2, D-64289 Darmstadt, Germany
	}
        \author{G.~Rainovski}
	\affiliation{Faculty of Physics, University of Sofia St. Kliment Ohridski, 5 James Bourchier blvd., BG-1164 Sofia, Bulgaria}
	\author{P.~R.~John}
    \affiliation{
		Institute for Nuclear Physics, Dept. of Physics, Technische Universität Darmstadt, Schlossgartenstraße 9, D-64289 Darmstadt, Germany
	}
	\author{M.~Beckers}
	\author{D.~Bittner}
    \author{A.~Blazhev}
    \affiliation{
		Institute for Nuclear Physics, Universität zu Köln, Zülpicher Straße 77, D-50937 Cologne, Germany
	}
    \author{A.~Esmaylzadeh}
	\author{C.~Fransen}
    \author{J.~Garbe}
	\author{L.~Gerhard}
	\author{K.~Geusen}
    \affiliation{
		Institute for Nuclear Physics, Universität zu Köln, Zülpicher Straße 77, D-50937 Cologne, Germany
	}
    \author{K.~Gladnishki}
    \affiliation{Faculty of Physics, University of Sofia St. Kliment Ohridski, 5 James Bourchier blvd., BG-1164 Sofia, Bulgaria}
	\author{A.~Goldkuhle}
    \affiliation{
		Institute for Nuclear Physics, Universität zu Köln, Zülpicher Straße 77, D-50937 Cologne, Germany
	}
	\author{K.~E.~Ide}
    \affiliation{
		Institute for Nuclear Physics, Dept. of Physics, Technische Universität Darmstadt, Schlossgartenstraße 9, D-64289 Darmstadt, Germany
	}
    \author{J.~Jolie}
	\affiliation{
		Institute for Nuclear Physics, Universität zu Köln, Zülpicher Straße 77, D-50937 Cologne, Germany
	}
    \author{V.~Karayonchev}
	\affiliation{
		Institute for Nuclear Physics, Universität zu Köln, Zülpicher Straße 77, D-50937 Cologne, Germany
	}
	\affiliation{Tri University Meson Facility TRIUMF, 4004 Wesbrook Mall, Vancouver, BC V6T 2A3, Canada}
	\author{R.~Kern}
    \affiliation{
		Institute for Nuclear Physics, Dept. of Physics, Technische Universität Darmstadt, Schlossgartenstraße 9, D-64289 Darmstadt, Germany
	}
    \author{E.~Kleis}
	\author{L.~Klöckner}
    \affiliation{
		Institute for Nuclear Physics, Universität zu Köln, Zülpicher Straße 77, D-50937 Cologne, Germany
	}
    \author{D.~Kocheva}
	\affiliation{Faculty of Physics, University of Sofia St. Kliment Ohridski, 5 James Bourchier blvd., BG-1164 Sofia, Bulgaria}
    \author{M.~Ley}
    \affiliation{
		Institute for Nuclear Physics, Universität zu Köln, Zülpicher Straße 77, D-50937 Cologne, Germany
	}
    \author{H.~Mayr}
	\author{N.~Pietralla}
    \affiliation{
		Institute for Nuclear Physics, Dept. of Physics, Technische Universität Darmstadt, Schlossgartenstraße 9, D-64289 Darmstadt, Germany
	}
    \author{F.~von Spee}
    \author{M.~Steffan}
    \affiliation{
		Institute for Nuclear Physics, Universität zu Köln, Zülpicher Straße 77, D-50937 Cologne, Germany
	}
    \author{T.~Stetz}
    \author{J.~Wiederhold}
    \affiliation{
		Institute for Nuclear Physics, Dept. of Physics, Technische Universität Darmstadt, Schlossgartenstraße 9, D-64289 Darmstadt, Germany
	}

	\date{\today}
	
	\begin{abstract}

    The lifetime of the $2^+_1$ state of \ce{^{210}Pb} was measured in the \ce{^{208}Pb}(\ce{^{18}O}, \ce{^{16}O})\ce{^{210}Pb} two-neutron transfer reaction by $\gamma$-ray spectroscopy employing the recoil-distance Doppler-shift method.  The extracted absolute \bdown value of $\asymUnc{119}{9}{8}{e^2\femto\metre^4}$ is consistent with previously reported measurements, but with significantly improved precision. The available experimental data for the $2^+_1$--$4^+_1$--$6^+_1$--$8^+_1$ multiplet are compared with shell-model calculations based on the well-established Kuo–Herling interaction. The new \bdown value agrees well with the shell-model prediction, providing evidence that the properties of the $2^+_1$ and $8^+_1$ states of \ce{^{210}Pb} can be consistently described together within the nuclear shell-model framework.
    
        
	\end{abstract}
	
	\maketitle
	
	
	\section{Introduction}

    The Nuclear Shell Model provides a foundational framework for nuclear structure physics, notably explaining the emergence of magic numbers. The magic numbers arise from significant energy gaps between nucleonic orbitals determined by the shape of the nuclear potential and the spin-orbit coupling~\cite{Mayer50,Jensen55}. In conjunction with pairing correlations, the shell model is instrumental in interpreting the low-energy spectra of semi-magic nuclei. In these systems, low-energy excitations with total angular momentum $J > 0$, involving two or more valence nucleons occupying a single high-$j$ orbital, typically originate from the angular momentum recoupling of unpaired nucleons. Such configurations form multiplets of states characterized by the number of unpaired nucleons, the quantum number $\nu$ known as seniority~\cite{Shalit63, Talmi71}. The generalized seniority scheme~\cite{Talmi71} is, in fact, an effective truncation of full shell-model calculations.

    Shell-model calculations describe the experimental properties of excited nuclear states particularly well for nuclei near shell closures, where the valence spaces are computationally tractable using standard shell-model techniques. Modern approaches, such as Monte Carlo methods~\citep{Koonin97, Otsuka01}, even allow access to nuclei approaching mid-shell regions, which involve a large number of valence nucleons. Nevertheless, nuclei with only a few valence nucleons, particularly those with two identical nucleons, are of special importance. These nuclei serve as benchmarks for validating microscopic nuclear interactions and for constraining key effective parameters such as nucleon effective charges and $g$-factors. Semi-magic nuclei with just two valence particles are also central to the seniority scheme. These nuclei serve as reference points, providing the properties of the fundamental $j^2$ configurations, such as the energies of states with $\nu = 2$, as well as the absolute $E2$ transition strengths for seniority-changing transitions ($\Delta \nu = 2$, $2^+_1 \rightarrow 0^+_1$) and seniority-conserving transitions ($\Delta \nu = 0$) within the $\nu = 2$ multiplet~\citep{Talmi71}.

    The nucleus \ce{^{212}Po} provides a representative example of how the properties of semi-magic nuclei with two valence particles can influence the structure of open-shell nuclei. \ce{^{212}Po} contains two valence protons and two valence neutrons relative to the doubly magic \ce{^{208}Pb} core. It has been demonstrated~\cite{Kocheva17a} that the experimentally determined absolute transition strength for the $2^+_1 \rightarrow 0^+_1$ transition of \ce{^{212}Po} is smaller by a factor of 1.5 to 2.5 compared to values predicted by shell-model calculations. This discrepancy has been linked to the properties of the simpler two-valence particle-systems \ce{^{210}Pb} and \ce{^{210}Po}. In both nuclei, the energies of the yrast $2^+$, $4^+$, $6^+$, and $8^+$ states exhibit seniority-like behaviour, characterized by decreasing energy spacings between adjacent levels as the spin increases. While these energy patterns are well reproduced by both single-$j$ and full shell-model calculations, none of the shell-model approaches employed have consistently described the corresponding \bdown transition strengths~\cite{Kocheva17a}. In the case of \ce{^{210}Po}, this problem is well known~\cite{Caurier03} and has raised concerns regarding the accuracy of the adopted $B(E2; 2^+_1 \rightarrow 0^+_1)$ value of \SI{42 \pm 9}{e^2\femto\metre^4}~\cite{Ellegaard73}. As a result, a remeasurement of the lifetime of the $2^+_1$ state of \ce{^{210}Po} was carried out~\cite{ref_kocheva_210po}. The newly determined value of $B(E2)_\text{exp} = \SI{136 \pm 21}{e^2\femto\metre^4}$~\cite{ref_kocheva_210po} has mitigated the disagreement between shell-model predictions of $B(E2)_\text{SM} = \SI{135}{e^2\femto\metre^4}$ and experimental data for the $2^+_1 \rightarrow 0^+_1$ transition of \ce{^{210}Po}, although significant discrepancies still remain for the other yrast transitions~\cite{Kocheva17a,Gerathy21,Stuchbery22}.

    It is noteworthy that the $B(E2; 2^+_1 \rightarrow 0^+_1)$ value of \ce{^{210}Pb} was determined using the same experimental technique as that employed to extract the earlier \bdown value of \ce{^{210}Po}~\cite{Ellegaard73} which was proven incorrect, namely, cross-section measurements from triton scattering experiments~\cite{ref_ellegard}.
    Even though the experimental \bdown value of \ce{^{210}Pb} of \SI{105 \pm 30}{e^2\femto\metre^4}~\cite{ref_ellegard} and shell-model predictions of \SI{109}{e^2\femto\metre^4} agree, a precise experimental value measured using a model-independent technique is necessary. The recently proposed theoretical framework for describing the structure of odd-even semi-magic and open-shell nuclei within a single-$j$ approximation with state-dependent effective charges~\cite{Kara19,Ley23} requires accurate \bdown values in the neighbouring nuclei with $j^2$ configurations. The experimental uncertainty associated with the $B(E2; 2^+_1 \rightarrow 0^+_1)$ value of \ce{^{210}Pb} exceeds \SI{30}{\%}, which prevents the applicability of this approach for nuclei above \ce{^{208}Pb}. This limitation, together with the need to verify the adopted $B(E2; 2^+_1 \rightarrow 0^+_1)$ value of \ce{^{210}Pb}, has motivated the present measurement of the lifetime of the $2^+_1$ state. The recoil-distance Doppler-shift (RDDS) method~\cite{RDDS, ref_plungerbible} was applied following a two-neutron transfer reaction to enable an accurate, model-independent determination of the $E2$ transition strength with higher precision than available in the literature.

	\section{Experiment}
	
	Low-lying excited states of \ce{^{210}Pb} were populated using the \ce{^{208}Pb}(\ce{^{18}O}, \ce{^{16}O})\ce{^{210}Pb} two-neutron transfer reaction. The \ce{^{18}O} beam was accelerated up to \SI{85}{\mega\electronvolt} by the \SI{10}{\mega\volt} FN-Tandem accelerator at the University of Cologne. The beam impinged on a target consisting of \SI{0.8}{\milli\gram\per\centi\metre^2} \ce{^{208}Pb} of \SI{99.14}{\%} enrichment, which was evaporated on a \SI{0.4}{\milli\gram\per\centi\metre^2} \ce{Mg} foil. Another \SI{1.7}{\milli\gram\per\centi\metre^2} \ce{Mg} foil was placed parallel to the \ce{^{208}Pb} target and used to stop the recoiling nuclei. The target and stopper were mounted inside the Cologne plunger device~\cite{ref_plungerbible}.
    
    \begin{figure}
		\includegraphics{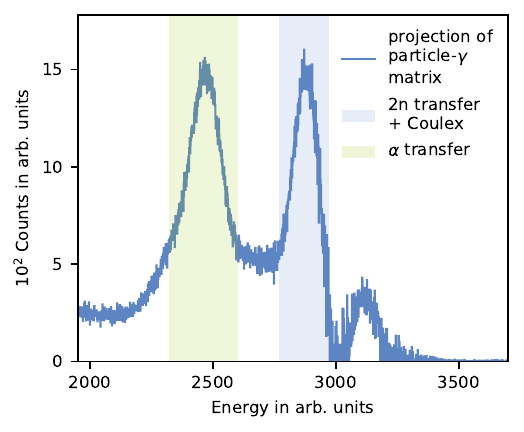}
		\caption{\label{part_spec} Projection of a particle-$\gamma$ matrix showing the relevant region for selecting the reaction products of interest. The \textit{green} region indicates the peak originating from back-scattered \ce{^{14}C} nuclei associated with the $\alpha$ transfer whereas the \textit{blue} region includes both events from \ce{^{16}O} and \ce{^{18}O} referring to two-neutron transfer and Coulomb excitation, respectively.}
    \end{figure}
    

    The plunger device is used to vary and keep the distance between target and stopper and, thus, $\gamma$-ray energy spectra were taken at eleven target-to-stopper distances, i.e. \SI{13}{\micro\metre}, \SI{22}{\micro\metre}, \SI{37}{\micro\metre}, \SI{47}{\micro\metre}, \SI{82}{\micro\metre}, \SI{142}{\micro\metre}, \SI{212}{\micro\metre}, \SI{282}{\micro\metre}, \SI{382}{\micro\metre}, \SI{512}{\micro\metre} and \SI{912}{\micro\metre}. The distances are determined with respect to the point of electrical contact of the target and stopper foil using the capacitance method with a distance offset of \SI{12 \pm 1}{\micro\metre}~\cite{ref_plungerbible, ref_capacitance_method}. The $\gamma$~rays emitted from the de-exciting nuclei were detected by eleven high-purity germanium (HPGe) detectors which were mounted in two rings. The five forward angle detectors were mounted at \SI{45}{\degree} and the six backward angle detectors at \SI{142}{\degree} with respect to the beam axis for the sensitive detection of Doppler shifts.
    
	The particles which were back-scattered in the reaction, i.e. the beam-like \ce{^{16}O} and \ce{^{18}O}, were detected by six solar cells placed at backward angles. The solar cells facilitate the creation of particle-gated spectra in order to separate the different occurring nuclear reactions and select the two-neutron transfer. Fig.~\ref{part_spec} shows an exemplary particle spectrum. The two most prominent peaks include events from $\alpha$ transfer (green) or Coulomb excitation and two-neutron transfer (blue). The $\alpha$ transfer is associated to the detection of \ce{^{14}C} nuclei and produces \ce{^{212}Po} which was analyzed in \cite{ref_vasil_212po}. Coulomb excitation and two-neutron transfer are related to \ce{^{18}O} and \ce{^{16}O}, respectively, and cannot be separated with the solar cells.
    
    \begin{figure}
		\includegraphics{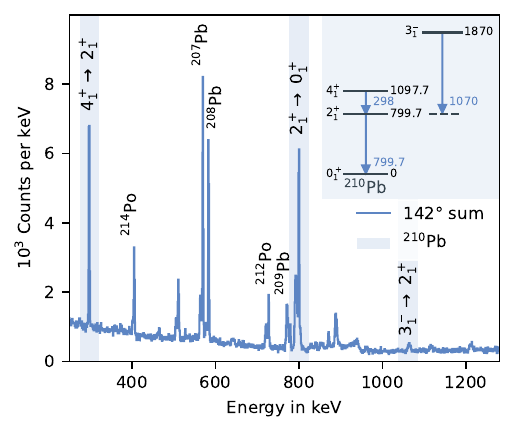}
		\caption{\label{pb_transitions_spec} Particle-gated $\gamma$-ray singles spectrum at \SI{142}{\degree} with all distances summed up. The occurring transitions of \ce{^{210}Pb} relevant for the analysis are indicated (\textit{blue}) as well as lines originating from \ce{^{207}Pb}, \ce{^{208}Pb}, \ce{^{209}Pb}, \ce{^{212}Po} and \ce{^{214}Po}. The inset shows a part of the level scheme of \ce{^{210}Pb} with the observed transitions.}
    \end{figure}
    
    The $\gamma$-ray singles spectrum at the backward angle of \SI{142}{\degree} obtained by particle-gating on the two-neutron transfer reaction and summing up all distances is shown in Fig.~\ref{pb_transitions_spec}. The transitions of \ce{^{210}Pb} which are relevant for the analysis are indicated as well as further occurring lines of transitions of other nuclei.
    
    The velocity of the recoiling \ce{^{210}Pb} nuclei is determined by exploiting the Doppler-shifted energy of the $3^-_1 \rightarrow 2^+_1$ transition. The more intense transitions were not chosen for the velocity determination, as the peaks of the $2^+_1 \rightarrow 0^+_1$ transition are subject to contaminations and the $4^+_1 \rightarrow 2^+_1$ transition barely develops a Doppler-shifted component due to the comparably large lifetime of the $4^+_1$ state. The velocity of the \ce{^{210}Pb} nuclei amounts to $v/c = \SI{1.2 \pm 0.1}{\%}$.
	
	\section{Analysis}
		
    The $2^+_1$ state of \ce{^{210}Pb} was analyzed applying the recoil-distance Doppler-shift (RDDS) method and its lifetime was determined using the differential decay curve method (DDCM). Both methods are explained in detail in \linebreak Ref. \cite{ref_plungerbible}. In an RDDS experiment the nucleus of interest is produced in a nuclear reaction and excited states are populated. The nuclei recoil out of the target foil in the direction of a stopper material. The decay of the excited states can take place when the nuclei are still in flight yielding a Doppler-shifted peak in the energy spectra whereas a de-excitation in the stopper results in a peak at the energy difference of the corresponding levels of the transition.
    
    
    The lifetime of the $2^+_1$ state of \ce{^{210}Pb} can directly be obtained from the distance-dependent ratios of the intensities of the shifted and unshifted components if the state is only directly populated in the nuclear reaction and no feeding from higher-lying states is present. However, the $4^+_1 \rightarrow 2^+_1$ transition of \ce{^{210}Pb} is visible in the spectra, such that a correction for feeding is required and was performed. As the statistics are not sufficient to create particle-gated $\gamma$-$\gamma$ matrices, the feeding intensity was explicitly subtracted from the intensity of the unshifted component of the $2^+_1 \rightarrow 0^+_1$ transition. Due to the competing fusion-evaporation channel yielding transitions at similar energies as the $2^+_1 \rightarrow 0^+_1$ transition of \ce{^{210}Pb}, using only $\gamma$-gated spectra for the analysis is not feasible. The only other observed feeder is the $3^-_1 \rightarrow 2^+_1$ transition which is visible only in the sum spectra and its lifetime is short in comparison to the one of the $2^+_1$ state. Hence and as the two-neutron transfer reaction favours the population of lower-lying states, further unobserved side feeding is considered to be negligible.

    \begin{figure}
		\includegraphics{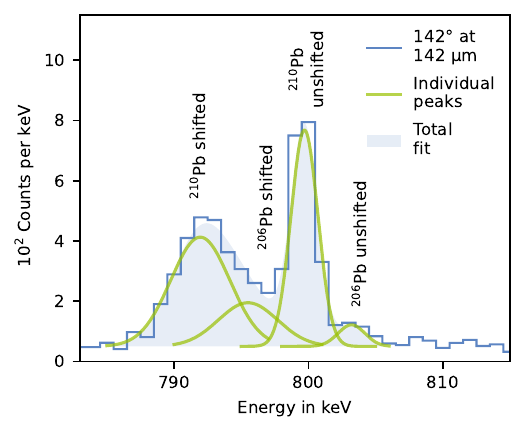}
		\caption{\label{pb_206pbcont_spec} Particle-gated $\gamma$-ray singles spectrum at \SI{142}{\degree} at \SI{142}{\micro\metre}. The shifted and unshifted components of the \linebreak \ce{{2}^+_1} $\rightarrow$ \ce{{0}^+_1} transitions of \ce{^{206}Pb} and \ce{^{210}Pb} are indicated as well as their corresponding fits (\textit{green}).}
    \end{figure}

    Further, the $2^+_1 \rightarrow 0^+_1$ transition of \ce{^{210}Pb} is subject to contaminations that need to be corrected before the lifetime of the $2^+_1$ state can be determined. The $\gamma$-ray energy spectrum in Fig.~\ref{pb_206pbcont_spec} shows the shifted and unshifted component of the $2^+_1 \rightarrow 0^+_1$ transition of \ce{^{210}Pb} measured at the backward angle of \SI{142}{\degree} with a contaminant identified as the $2^+_1 \rightarrow 0^+_1$ transition of \ce{^{206}Pb}. Despite the high enrichment of the target material, small impurities and their Coulomb excitation explain the occurrence of the transition of \ce{^{206}Pb} in the spectra. The lifetime of the $2^+_1$ state of \ce{^{206}Pb} is known to be \SI{11.8 \pm 0.1}{\pico\second} \cite{nds206}. It is in the same order of magnitude as the lifetime of the $2^+_1$ state of \ce{^{210}Pb}. Therefore, the $2^+_1 \rightarrow 0^+_1$ transition of \ce{^{206}Pb} also develops a shifted component. The shifted component of \ce{^{206}Pb} overlaps with both components of the $2^+_1 \rightarrow 0^+_1$ transition of \ce{^{210}Pb}. In order to correctly account for the \ce{^{206}Pb} contamination, an effective lifetime of the $2^+_1$ state of \ce{^{206}Pb} was estimated from the decay behaviour of its unshifted component at a detector angle of \SI{142}{\degree}. Using this effective lifetime, the peak area of the shifted component of the $2^+_1 \rightarrow 0^+_1$ transition of \ce{^{206}Pb} is calculated for the different plunger distances. The \ce{^{206}Pb} peak areas are fixed during fitting as shown in Fig.~\ref{pb_206pbcont_spec}.

    \begin{figure}
		\includegraphics{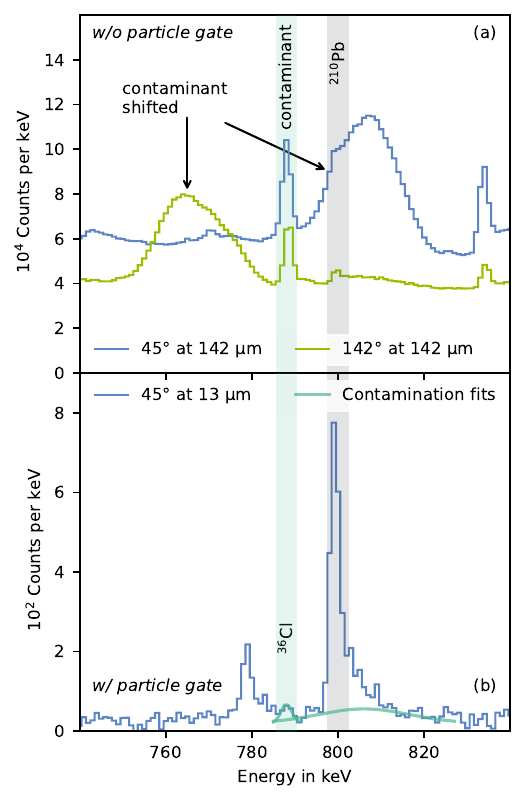}
		\caption{\label{pb_cl_cont_spec} (a) $\gamma$-ray singles spectrum at \SI{45}{\degree} (\textit{blue}) and \SI{142}{\degree} (\textit{green}) at \SI{142}{\micro\metre} without a particle gate. The unshifted components of the \ce{{2}^+_1} $\rightarrow$ \ce{{0}^+_1} transition of \ce{^{210}Pb} (\textit{grey}) as well as the \ce{{3}^+_1} $\rightarrow$ \ce{{2}^+_1} transition of the \ce{^{36}Cl} contaminant (\textit{cyan}) are indicated.\\
		(b) Particle-gated $\gamma$-ray singles spectrum at \SI{45}{\degree} at \SI{13}{\micro\metre}. The fits of the shifted and unshifted components of the \ce{^{36}Cl} contaminant (\textit{cyan}) are indicated. The energies, peak widths and areas of the contaminant were determined with the ungated spectra and kept fixed for the correction.}
    \end{figure}
    
    The $\gamma$-ray singles spectra created without applying a particle gate reveal another contaminant. The spectra are shown in Fig.~\ref{pb_cl_cont_spec} (a) at an intermediate distance of \SI{142}{\micro\metre}. At \SI{788}{\kilo\electronvolt} the unshifted component of the contaminant is visible. The comparison of the two detector ring angles indicates the development of a shifted component of the contaminant. At the forward angle of \SI{45}{\degree} the shifted component overlaps with both components of the $2^+_1 \rightarrow 0^+_1$ transition of \ce{^{210}Pb}. The contaminant could be identified as the $3^+_1 \rightarrow 2^+_1$ transition of \ce{^{36}Cl} which is produced in the fusion-evaporation channel following the reaction of the \ce{^{18}O} beam with Mg.

    \begin{figure}
	\includegraphics{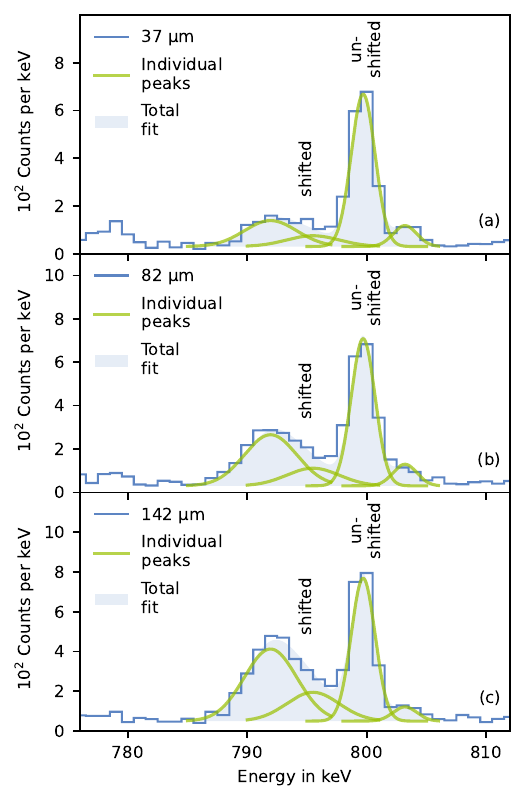}
	\caption{\label{pb_doppler_spec} Particle-gated $\gamma$-ray singles spectra at \SI{142}{\degree} at (a) \SI{37}{\micro\metre}, (b) \SI{82}{\micro\metre} and (c) \SI{142}{\micro\metre}. The fits of the \ce{{2}^+_1} $\rightarrow$ \ce{{0}^+_1} transition of both \ce{^{210}Pb} and the contaminant \ce{^{206}Pb} are depicted (\textit{green}). The evolution of the intensity ratio of the shifted and unshifted component is visible. The intensity of the shifted component increases with the distance while the unshifted component decreases.}
    \end{figure}
    
    A small fraction of the \ce{^{36}Cl} contaminant is left by the particle gate and the random background subtraction and remains in the gated spectrum shown in \linebreak Fig.~\ref{pb_cl_cont_spec} (b). At \SI{788}{\kilo\electronvolt} a peak associated with the unshifted component of the $3^+_1 \rightarrow 2^+_1$ transition of \ce{^{36}Cl} is visible. As the intensity ratio of the shifted and unshifted components of the contaminant does not change with the distance in the ungated spectra, meaning that most of the reaction producing \ce{^{36}Cl} occurs in the stopper, it is used to estimate the intensity of the shifted component in the gated spectra. The energy and width of the shifted component of the $3^+_1 \rightarrow 2^+_1$ transition of \ce{^{36}Cl} are determined with the ungated spectra as well in order to fix them in the gated spectra. The fits of the contaminant peaks are indicated in Fig.~\ref{pb_cl_cont_spec} (b) and allow for the accurate determination of the peak areas of the shifted and unshifted components of the $2^+_1 \rightarrow 0^+_1$ transition of \ce{^{210}Pb}.
    
    \begin{figure}
	\includegraphics{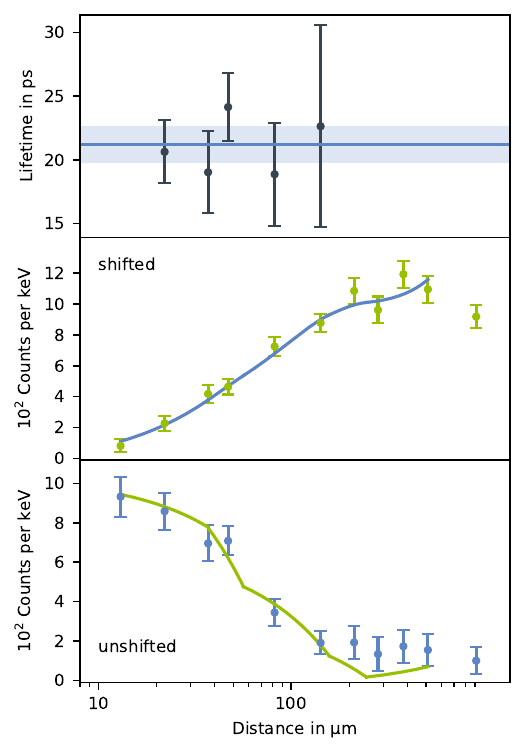}
	\caption{\label{napa_bw} Exemplary \texttt{napatau} fits obtained with the DDCM analysis at \SI{142}{\degree}. The \textit{top} plot shows the values of the distance-wise determined lifetimes with their mean value and uncertainty. The \textit{centre} and \textit{bottom} plot illustrate the second degree spline fits of the intensities of the shifted and unshifted component. The distances for which the intensities remain constant are not sensitive for the lifetime measurement and are, thus, omitted from the mean value determination.}
    \end{figure}
    
    \begin{figure}[t]
		\includegraphics{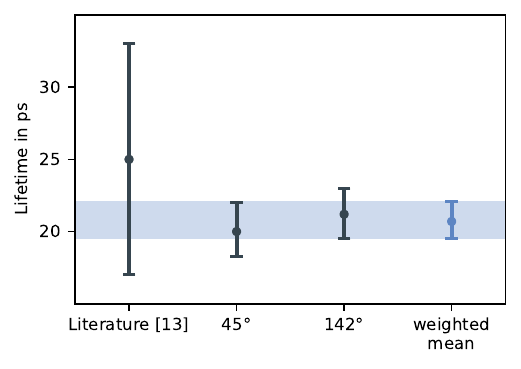}
		\caption{\label{lifetimes} Comparison between the values for the lifetime of the \ce{{2}^+_1} $\rightarrow$ \ce{{0}^+_1} transition of \ce{^{210}Pb} from literature \cite{ref_ellegard} and this work showing the individual results at detector angles of \SI{45}{\degree} and \SI{142}{\degree} and the weighted mean (\textit{blue}).}
    \end{figure}
    
    The DDCM analysis determines the lifetime $\tau$ at every distance following
	\begin{equation} \label{ddcmeq}	
		\tau = \frac{I_\text{u}}{\frac{\text{d}}{\text{d}t} I_\text{s}} = \frac{I_\text{u}}{v \hspace{2pt} \frac{\text{d}}{\text{d}x} I_\text{s}}
	\end{equation}
    
	with the intensities of the shifted and unshifted components $I_s$ and $I_u$ and the velocity $v$. The evolution of the intensities of both components with the distance is depicted in Fig.~\ref{pb_doppler_spec} showing three exemplary distances. Due to the differential approach of the DDCM only relative distances are required which has the advantage of smaller uncertainties in comparison to absolute distances. The initial distance dependence of the intensities is converted into a time dependence with the known velocity~$v$ of the \ce{^{210}Pb} nuclei. The mean lifetime of all distances is determined with a $\chi^2$ fit implemented in the program \texttt{napatau}~\cite{ref_napatau} and exemplary results are shown in Fig.~\ref{napa_bw}.
    
    The \textit{top} part displays the distance-wise determined lifetime values with their mean value as the blue line and corresponding uncertainty as the blue band. The \textit{centre} and \textit{bottom} parts illustrate the second order spline fits of the shifted and unshifted components that describe the evolution of the intensities with the plunger distance.
    
    The DDCM analysis was performed for both detector ring angles separately yielding \asymUnc{20.0}{2.0}{1.7}{\pico\second} at the forward angle of \SI{45}{\degree} and \asymUnc{21.2}{1.8}{1.7}{\pico\second} at the backward angle of \SI{142}{\degree}. The weighted mean value results in a lifetime of $\tau = \asymUnc{20.7}{1.4}{1.2}{\pico\second}$ which agrees with the literature value within its uncertainty. Fig.~\ref{lifetimes} shows a comparison between the literature value for the lifetime of the $2^+_1$~state and the results of this work and illustrates the improved precision. The corresponding quadrupole transition strength was calculated using a conversion coefficient of $\alpha = \num{0.01040 \pm 0.00001}$ \cite{bricc} and amounts to 
	\begin{equation*}
		B(E2; 2^+_1 \rightarrow 0^+_1) = \asymUnc{119}{9}{8}{e^2\femto\metre^4} = \asymUnc{1.61}{0.12}{0.10}{\weisskopfunit}
	\end{equation*}






    \section{Discussion}
    
    The available experimental data on the properties of the yrast states of \ce{^{210}Pb} are summarized in Tables~\ref{210Pb} and~\ref{mq}. The newly determined \bdown value for the $2^+_1~\rightarrow~0^+_1$~transition of \asymUnc{119}{9}{8}{e^2\femto\metre^4} is in good agreement with the previously reported value of \SI{105 \pm 30}{e^2\femto\metre^4} (cf.~Table~\ref{210Pb}), but with significantly improved precision. This represents the main experimental result of the present study.

    \begin{table}[t]
    \centering
    \caption{Comparison between the experimental and calculated (see text for details) properties of the yrast states of \ce{^{210}Pb}.}
    \label{210Pb}
    \begin{ruledtabular}
    \begin{tabular}{cccccc}
    \multicolumn{1}{c}{$J^\pi_i$} & \multicolumn{2}{c}{$E_x$ in \si{\mega\electronvolt}} & \multicolumn{1}{c}{$J^\pi_f$} &\multicolumn{2}{c}{$B(E2;J_i \rightarrow J_f)$ in \si{e^2\femto\metre^4}} \\ 
    \cline{2-3}\cline{5-6}
    \noalign{\vskip 1pt}
    & Expt\footnotemark[1] & SM & & Expt & SM\footnotemark[2] \\ \noalign{\vskip 1pt}\hline
    \noalign{\vskip 2pt}
    $2^+_1$ & \num{0.800 \pm 0.001} & 0.837 & $0^+_1$ & $119^{+9}_{-8}$\footnotemark[3] & 126(15) \\
                 &          &          &               & 105(30)\footnotemark[4] &             \\ 
    $4^+_1$ & \num{1.098 \pm 0.001} & 1.099 & $2^+_1$ & 356(67)\footnotemark[5] & 168(20) \\ 
    $6^+_1$ & \num{1.195 \pm 0.004} & 1.191 & $4^+_1$ & 87(10)\footnotemark[6]  & 118(13) \\
                 &          &          &               & 103(35)\footnotemark[7]  &        \\
                 &          &          &               & 156(6)\footnotemark[8]  &        \\ 
    $8^+_1$ & \num{1.278 \pm 0.005} & 1.234 & $6^+_1$ & 49(5)\footnotemark[9]  &  49(5)
    \end{tabular}
    \footnotetext[1]{The experimental excitation energies are taken from Ref.~\cite{nds14}.}
    \footnotetext[2]{With $e_\nu = \num{0.88 \pm 0.05}\,e$, for details see the text.}
    \footnotetext[3]{From the present work.}
    \footnotetext[4]{From Ref.~\cite{ref_ellegard}.}
    \footnotetext[5]{From the lifetime of the $4^+_1$ state reported in Ref.~\cite{Weinzierl64}.}
    \footnotetext[6]{From the lifetime of the $6^+_1$ state of \ce{^{210}Pb} reported in Ref.~\cite{Broda18}.}
    \footnotetext[7]{From the lifetime of the $6^+_1$ state of \ce{^{210}Pb} determined by gating on a direct feeding transition reported in Ref.~\cite{Broda18}.}
    \footnotetext[8]{From the lifetime of the $6^+_1$ state of \ce{^{210}Pb} reported in Ref.~\cite{Decman83}.}
    \footnotetext[9]{From the lifetime of the $8^+_1$ state of \ce{^{210}Pb} reported in Ref.~\cite{Decman83} and confirmed in Ref.~\cite{Broda18}.}
    \end{ruledtabular}
    \end{table}

    To assess how the new $B(E2; 2^+_1 \rightarrow 0^+_1)$ value fits within the widely accepted shell-model description of the yrast states of \ce{^{210}Pb}~\cite{Kocheva17a,Broda18}, we performed shell-model calculations using the code NuShellX~\cite{NuShellX} and the well-established Kuo-Herling interaction, as modified by Warburton and Brown (denoted KHPE in the NuShellX package)~\cite{khpe}. The neutron valence space encompassed the full $N = 126$–$184$ major shell, including the $1g_{9/2}$, $0i_{11/2}$, $1g_{7/2}$, $2d_{5/2}$, $2d_{3/2}$, $3s_{1/2}$, and $0j_{15/2}$ orbitals. The standard effective gyromagnetic factors $g_{\pi}^{l} = 1.107$, $g_{\nu}^{l} = -0.065$, $g_{\pi}^{s} = 3.234$, and $g_{\nu}^{s} = -2.083$ were adopted from Ref.~\cite{eg}, while the $g_{p,\nu}$-factor related to the tensor component of the $M1$ operator was fixed to \num{-0.473} to reproduce the magnetic moment of the $9/2^+$ ground state of \ce{^{209}Pb} \cite{Anselment86}.
    
    The radial part of the wave functions was calculated using a harmonic oscillator potential with $\hbar\omega~=~\SI{6.863}{\mega\electronvolt}$. As suggested in Refs.~\cite{Broda18,Decman83}, the neutron effective charge was fixed to reproduce the experimental $B(E2; 8^+_1 \rightarrow 6^+_1)$ value, resulting in $e_\nu~=~\num{0.88 \pm 0.05}\,e$. The main results from these shell-model calculations are presented under the columns labeled SM in Tables~\ref{210Pb} and~\ref{mq}.

    The calculated level energies are in good agreement with the experimental values for the $2^+_1$, $4^+_1$, $6^+_1$, and $8^+_1$ states, with the largest discrepancy of \SI{44}{\kilo\electronvolt} observed for the $8^+_1$ state (cf. Table~\ref{210Pb}). The wave functions of the yrast states of \ce{^{210}Pb} indicate that they are nearly pure seniority-2 states, with the $\nu(g_{9/2})^2$ configuration accounting for \SI{94.5}{\%} of the $2^+_1$ state's structure and increasing to \SI{98.9}{\%} for the $8^+_1$ state, as previously noted in Ref.~\cite{Broda18}. The next most significant components are $\nu(g_{9/2})^1\nu(d_{5/2})^1$, $\nu(g_{9/2})^1\nu(d_{3/2})^1$, and $\nu(j_{15/2})^2$, each contributing less than \SI{1}{\%} to the total wave functions.
    
    Both, experiment and shell-model calculations yield nearly constant $g$-factors for the $8^+_1$ and $6^+_1$ states. The shell-model calculations reproduce the experimental magnetic moments quantitatively (cf. Table~\ref{mq}). This agreement provides strong indication that these levels are indeed built on identical configurations characterized by good seniority.


    The newly measured $B(E2; 2^+_1 \rightarrow 0^+_1)$ value agrees with the shell-model prediction obtained using an effective neutron electric charge of $\num{0.88 \pm 0.05}\,e$, which was fixed to reproduce the \bdown value of the $8^+_1 \rightarrow 6^+_1$ transition (cf. Table~\ref{210Pb}). This result stands in stark contrast to the case of \ce{^{210}Po}~\cite{Kocheva17a,ref_kocheva_210po}, and provides strong evidence that the properties of the $8^+_1$ and $2^+_1$ states of \ce{^{210}Pb} can be consistently described within the nuclear shell-model framework as being dominated by seniority $\nu = 2$.

    The calculated $B(E2; 6^+_1 \rightarrow 4^+_1)$ value of \SI{118 \pm 13}{e^2\femto\metre^4} differs markedly from the experimental value of \SI{156 \pm 6}{e^2\femto\metre^4} reported in Ref.~\cite{Decman83} and only slightly overestimates the experimental result of \SI{87 \pm 10}{e^2\femto\metre^4} given in Ref.~\cite{Broda18} which is calculated from the weighted average of two lifetime values for the $6^+_1$ state that are determined with different gates. The discrepancy between shell-model calculations and experimental results may suggest that the individual lifetime result of the $6^+_1$~state that was determined by gating on a direct feeding transition and reported as \SI{113 \pm 22}{\nano\second} in Ref.~\cite{Broda18}, corresponding to a \bdown value of \SI{103 \pm 35}{e^2\femto\metre^4}, is more accurate than the adopted weighted average of \SI{92 \pm 10}{\nano\second} reported in Ref.~\cite{Broda18}.

    \begin{table}[t]
    \centering
    \caption{Comparison between the experimental and calculated (see text for details) magnetic ($\mu$) and quadrupole ($Q$) moments of \ce{^{210}Pb} and \ce{^{209}Pb} isotopes.
    The experimental data is taken from Refs.~\cite{nds14,Anselment86}.}
    \label{mq}
    \begin{ruledtabular}
    \begin{tabular}{ccccc}    \multicolumn{1}{c}{} & \multicolumn{2}{c}{$\mu$ in $\si{\mu^2_N}$} & \multicolumn{2}{c}{$Q$ in \si{e\femto\metre^2}}  \\
    \cline{2-3}\cline{4-5}
    \noalign{\vskip 2pt}
                      & Expt & SM & Expt & SM \\ \hline
    \noalign{\vskip 2pt}
    \ce{^{210}Pb}  &                      &                  &           &      \\
    $6^+_1$      & -1.872(90)      &   -1.859     & --        &  -10(1) \\
    $8^+_1$      &  -2.496(64)     &   -2.501     & --        &  -40(2)  \\
    \noalign{\vskip 2pt}
    \hline
    \noalign{\vskip 2pt}
    \ce{^{209}Pb} &                        &                 &           & \vspace{2pt} \\
    $9/2^+_1$ & -1.4735(16) &  -1.4735 & -27(17) & -29(2) 
    \end{tabular}
    \end{ruledtabular}
    \end{table}

    The largest discrepancy between the calculated and experimental \bdown values is observed for the $4^+_1 \rightarrow 2^+_1$ transition, amounting to a factor of more than two. Given that the experimental value is based on an early lifetime measurement~\cite{Weinzierl64}, this significant deviation highlights the need for a renewed experimental determination of the lifetime of the $4^+_1$ state with higher precision and accuracy.

    \section{Summary}
    
    The lifetime of the $2^+_1$ state of \ce{^{210}Pb} was measured using the recoil-distance Doppler-shift method following a two-neutron transfer reaction. This measurement yielded a precise, model-independent value for the $E2$ transition strength of the $2^+_1 \rightarrow 0^+_1$ transition. The result is in good agreement with the previously reported value~\cite{ref_ellegard}, but with significantly improved precision. It is consistent with the shell-model prediction, supporting a coherent description of both the $2^+_1$ and $8^+_1$ states. Overall, the shell-model calculations provide a satisfactory account of the $2^+_1$--$4^+_1$--$6^+_1$--$8^+_1$ multiplet, as expected for states based on a nearly pure seniority-2 configuration. The enhanced precision of the new $B(E2; 2^+_1 \rightarrow 0^+_1)$ value strengthens its utility for modelling the structure of nuclei above Pb within a single-$j$ framework employing state-dependent effective charges. Nonetheless, the remaining discrepancy between the calculated and reported $E2$ transition strengths of the $4^+_1 \rightarrow 2^+_1$ transition underscores the need for an updated experimental determination of the lifetime of the $4^+_1$ state.
 
    \begin{acknowledgments}
    
    This work was supported by the German Federal Ministry of Education and Research (BMBF) under Grant Nos. 05P21RDCI2 and 05P24RD3 and by the Deutsche Forschungsgemeinschaft (DFG, German Research Foundation) as part of the Project\mbox{-}ID~264883531 – Research Training Group 2128 'Accelence' and Project\mbox{-}ID~499256822 – Research Training Group 2891 'Nuclear~Photonics'. This work was supported by DAAD under the partnership agreement between the University of Cologne and the University of Sofia and by the Bulgarian Ministry of Education and Science, within the National Roadmap for Research Infrastructures (object CERN). D.K. and K.G. acknowledge support by the European Union-NextGenerationEU, through the National Recovery and Resilience Plan of the Republic of Bulgaria, project No. BG-RRP-2.004-0008-C01.
    
    \end{acknowledgments} 
    
    \appendix
    
    \nocite{*}
    	
    \bibliography{210Pb_paper}
    	
    \end{document}